\documentclass[aps,pre,twocolumn,showpacs]{revtex4}
\usepackage{graphicx}
\usepackage{amssymb}
\usepackage{amsmath}

\begin{document}              
\widetext

\title{Condensation Phenomena in Nano-Pores - a Monte Carlo Study}
\thispagestyle{empty}

\author{Raja Paul{$^1$}, Heiko Rieger{$^1$}}

\affiliation{$^1$Theoretische Physik, Universit\"at des Saarlandes, 
66041 Saarbr\"ucken, Germany.}

\begin{abstract}
 The non-equilibrium dynamics of condensation phenomena in nano-pores
is studied via Monte Carlo simulation of a lattice gas model. Hysteretic
 behavior of the particle density as a function of the density of a 
 reservoir is obtained for various pore geometries in two and three
 dimensions. The shape of the hysteresis loops depend on the
 characteristics of the pore geometry. The evaporation of particles from
 a pore can be fitted to a stretched exponential decay of the
 particle density $\rho_f(t) \sim exp \left[ -(t/\tau)^\beta\right]$.
 Phase separation dynamics inside the pore is effectively described by a
 random walk of the non-wetting phases. Domain evolution is 
 significantly slowed down in presence of random wall-particle potential and
 gives rise to a temperature dependent growth exponent. On the other hand
 roughness of the pore wall only delays the onset of a pure domain growth.  
\end{abstract}

\pacs{64.60-i, 81.07-b, 47.55.Mh, 75.40.Gb}
\maketitle  

\section{introduction}

Adsorption and desorption isotherms of a gas condensed into nano pores
show hysteresis and have become very a useful tool for the classification of
porous materials~\cite{rouqerol,sing,gelb}. Due to the effect of surface
tension the formation of meniscii inside the pore is delayed on the
desorption branch, resulting in a non-vanishing hysteresis area of the sorption
isotherms. A theoretical analysis and detailed description of the early
research on this phenomenon is described in a review by
Everett~\cite{everett}.  Recent investigation from a point of view of
the stability of adsorbed multilayers~\cite{saam-cole} and
other~\cite{celestini, papadopoulou,evans}  
analytical and numerical approaches together with density functional
theory also   
proposed that the hysteresis phenomenon is an intrinsic property of the
phase transition in a single idealized pore and arises from the
existence of metastable states. 

Hysteresis in real materials is the collective phenomena
involving interconnected network of simple
pores~\cite{guyer96,mason88}. However, quite recently it has been shown by
numerical analysis~\cite{sarkisov} and
experiments~\cite{wallacher04,huber,page} 
that hysteresis can occur in isolated pores also. Moreover the shape of
the hysteresis loops are influenced by the characteristic
features of the pore geometry. In this paper, using extensive
Monte Carlo simulations, we intend to characterize the sorption
isotherms for different pore geometries and compare our results
directly to the experimental observations.  

Furthermore, we study the phase separation kinetics of a binary
liquid in nano-pores at low temperatures. Previous work~\cite{goh,dierker,lin94} 
showed that a binary liquid, unlike Ising-like complete phase
separation, does not separate into two phases completely. Instead the
adsorbed material forms many small domains far below the coexistence
region. An explanation for this behavior has been suggested on the
basis of random-field Ising model 
\cite{brochard,degennes,birgeneau}. On the
contrary we found in case of a
single-pore model with no randomness, confinement in a small pore
slows down domain growth in certain regions of the wetting phase
diagram~\cite{wetting-liu} and as a result macroscopic phase separation
is not observed on short time scales. Moreover the late stage domain
evolution, obtained from a two point correlation function 
$C(r,t)$  follows the same Lifshitz-Slyozov~\cite{lifshitz61} growth law of
$t^{1/3}$. Unlike earlier research~\cite{lee92} of domain evolution in
porous networks, we concentrate on a single pore which is more
applicable for systems with low porosity(Si
mesopores~\cite{wallacher04}). Moreover, randomness in terms of irregular pore 
structure or presence of impurity atoms in the pore wall, which is
inherent in real systems, may have drastic effect on the domain
growth. 

In this paper we focus on a single pore. An average
over many isolated pores corresponding to an assembly of
non-interconnected(unlike Vycor) pores can be experimentally
developed~\cite{wallacher04} out of a B-doped Si wafer via
electrochemical etching. The structure of 
the rest of the paper is as follows. First we overview, 
in section II, the theoretical model and the Monte Carlo technique that
we use in our simulation. In section III~A, we discuss the Hysteresis
phenomenon for different pore geometries in 2 and 3 dimensions and
discuss how one can differentiate the pore structure from the hysteresis
loops. In section III~B, we focus on the evaporation of particles from
the pore and discuss how the density in the pore reaches a new
equilibrium value. Sections III~C, D are entirely devoted to the domain
evolution in the pore environment where we propose a 
random-walk picture of the domain growth above a critical size
comparable to the pore diameter. Then we perform a quantitative study of
growth phenomenon for a simple pore as well as for pores with
defects. Finally section IV finishes the paper with a discussion.

\section{Simulation model}
A standard model for a binary liquid mixture, is the Ising
lattice-gas model with spin occupancy variables 
$\sigma_i$ = 0, 1 governed by the Hamiltonian:  

\begin{equation}
{\mathcal{H}} = -W_{pp}\sum_{bulk \langle i,j\rangle}{\sigma_i \sigma_j}
 - W_{wp}\sum_{i ~n.n. ~of ~wall}{\sigma_i} , 
\label{eq1}
\end{equation}
where $W_{pp}$ and $W_{wp}$ are the particle-particle and
wall-particle couplings respectively. We denote an
occupied site as a particle and an empty site as a vacancy. Experimental 
observations suggest that in case of a glass(Si) adsorbent, the pore wall has
a very strong affinity towards the adsorbed gases, which implies $W_{wp} 
> W_{pp}$. Most of our simulations, unless  
otherwise specified, are performed at a fixed ``wettability''
$W_{wp}/W_{pp}=1.5$. A change in this
value does not qualitatively affect our results as long as $W_{wp}
> W_{pp}$. The geometry of the pores is chosen to 
be a rectangle of size $L\times h$ in two dimension and parallelepiped of
size $L \times h\times h$ in three dimension with  $h \ll L$. Standard
conserved order parameter dynamics(Kawasaki) is
employed to study the diffusion and the domain growth
kinetics in the nano pores.  Nearest neighbor spins $\sigma_i$ and
$\sigma_j$ are exchanged with Metropolis acceptance probability   
\begin{eqnarray}
{P}(\{\sigma\}\longrightarrow \{\sigma'\})= 
 {\mathrm {min}} \left[
      \begin{array}{cll}
       1,~\frac{exp (-\beta \mathcal{H} \{\sigma'\})}
	{exp(-\beta \mathcal{H} \{\sigma\})}
      \end{array}
	  \right],
\label{eq2}
\end{eqnarray}
where $\sigma$ and $\sigma'$ represents the old and new spin
configuration respectively. To improve the speed of simulations we
always choose a particle with unit probability and define 1 MC step
containing a number of trial-updates that is equal to the number of 
particles present in the system.

\section{Numerical Results}
\subsection{Hysteresis}
\begin{figure}
 \includegraphics[width=\linewidth]
 {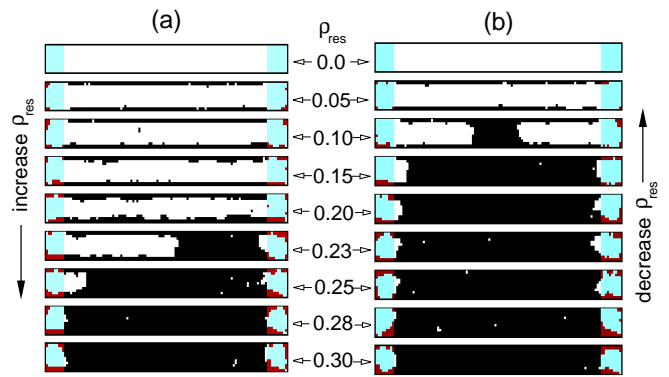}
 \caption{Snapshots of the hysteresis in a simple nano-pore of
 size $L=100$ and $h=11$ at temperature $T=0.3$. The shaded regions
 and the black spots on both sides of the pore are respectively the attached
 reservoirs and the particles in it. For each value of 
 $\rho_{res}$ the system is equilibrated up to $t=2^{22}$.
 Snapshots {\bf (a)} correspond to the adsorption and {\bf (b)}
 correspond to the desorption isotherms.
}
 \label{fig1}
\end{figure}

\begin{figure}
 \includegraphics[width=\linewidth]
{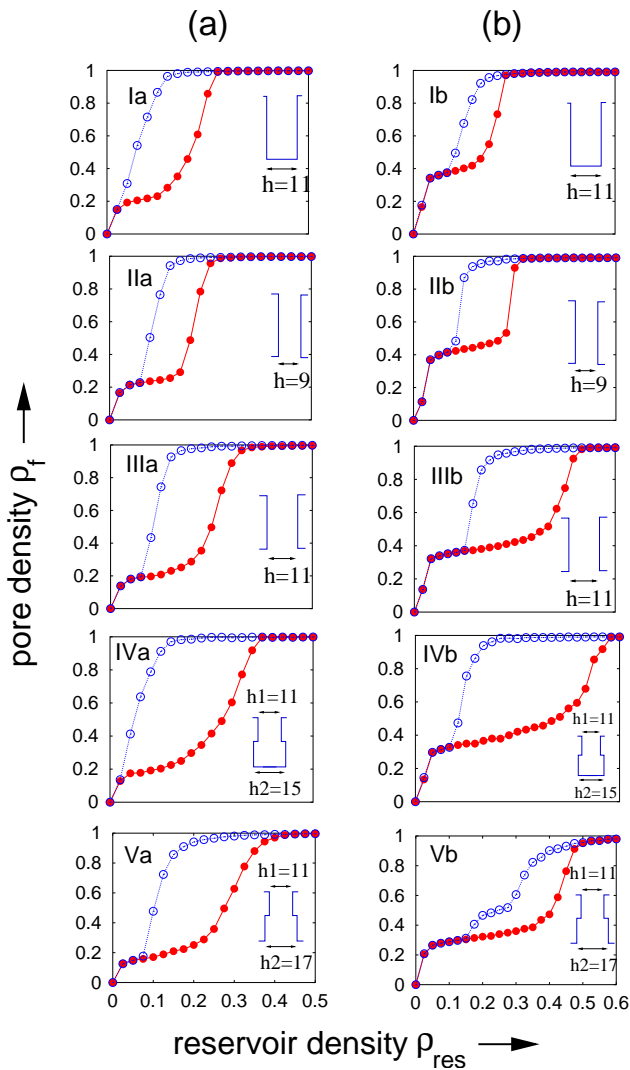}
\caption{ Adsorption (filled circles) and desorption(empty circles) isotherms
 for nano pores. {\bf (a):} Filling fraction $\rho_f$ of the pore as a function
 of the reservoir density $\rho_{res}$ in 2 dimension. The equilibration
 time $t_w$ for each data point $2^{22}$. Ia has one reservoir of length
 10 attached to the open end. IIa, IIIa and IVa has 2 reservoir attached
 on both sides. {\bf (b):} Hysteresis isotherms in 3 dimension. 
 Equilibration time $t_w = 2^{22}$, pore length $L = 50$ and reservoir
 length is 10. For each hysteresis isotherm data are averaged over 50 disorder
 samples. }
 \label{fig2}
\end{figure}

To study hysteresis in the adsorption/desorption process within a pore we
attach one or two reservoirs depending upon the geometry of the
pore. To avoid asymmetric 
diffusion into the pore, no periodic boundary condition is
applied between the reservoirs. The particle density in the reservoirs
is kept constant by adding(removing) particle at a randomly chosen
position in the reservoir as soon as a particle diffuses into the
pore(reservoir). A snapshot of the hysteresis phenomenon
for a simple pore geometry in $2d$ is shown in Fig.~\ref{fig1}. 
Initially, both the pore and the reservoirs are kept empty. The
density of particles in the reservoirs($\rho_{res}$) are then slowly
increased. Particles are immediately adsorbed and form a single layer
along the pore walls. At this stage the density in the 
pores($\rho_f$), rises sharply to a non-zero value which corresponds to
the first jump of adsorption isotherms(Fig.~\ref{fig2}). Then
$\rho_f$ forms a long plateau until two semicircular(or hemispherical in
$3d$) meniscii(a 
meniscus for one-end open pore) are formed somewhere in the middle of the 
pore at a high reservoir density and move apart from each other to fill the entire 
pore; a second jump in $\rho_f$ is observed. We call this
complete pore filling and further increase in $\rho_{res}$ does not
change $\rho_f$. 

Next, we slowly decrease the reservoir density and record the
corresponding filling fraction $\rho_f$ after the system is
equilibrated. On desorption, however at a much smaller reservoir density
$\rho_{res}$, we see the semicircular meniscii to form on the open ends
of the pore, followed by a sharp fall in the pore density $\rho_f$, as
the meniscii approaches towards each other. Finally the desorption curve
follows the same path as that of the adsorption.  Since the diffusion
occurs very slowly into the pores, a long waiting time $t_w$ is 
needed to equilibrate the whole system for each value of $\rho_{res}$. 

Our simulation results are shown in Fig.~\ref{fig2} where column
{\bf(a)} and {\bf(b)} shows the hysteresis in 2 and 3 dimensions
respectively. The simulations are performed with systems of length $L
=$ 100 and  50 at temperatures $T$ = 0.3, 0.6 in $2d$ and $3d$
respectively. Reservoir 
lengths are kept fixed at $L_r = 10$. Ink-bottle pores are equally
divided into two parts; one with the narrow tube of diameter $h_1$ and the
other is relatively wider sack with diameter $h_2$. To make sure that pore
density $\rho_f$ does not change any more, each data point of the
sorption-isotherms is equilibrated for $t_w=2^{22}$ time steps. Finally
for each hysteresis loop data are averaged over 50 ensembles.  It is
evident from the sorption branches of II-IIIa,b, that they are much steeper
than in Ia,b. This is due the formation of two meniscii on both sides of 
the pore, which accelerate the filling and emptying procedure. Such
characteristic feature distinguishes between a one-end open and both-end open
pore. Further, an increase in pore diameter requires a higher
reservoir density to initiate the formation of meniscii, which
effectively delays the onset of adsorption saturation. As a result the the loop
area for bigger pore is also increased as shown in IIa,b and
IIIa,b. This features can be used to compare pores of different sizes. 

In experiments~\cite{wallacher04} one observes a two step decrease of the
desorption isotherms corresponding to ink-bottle pores. Our data, as
shown in figure Vb, for a both-end open pore in $3d$ of similar
geometry, agrees quite well with the experimental predictions. The two
step decrease of the desorption isotherm arises from the sequential
emptying of the different sections(of diameter $h_1$ and
$h_2$) of the pore. The heights of the two steps suggest the wider
section of the pore emptied earlier than the smaller
section. However, ink-bottle pores in figure IVa,b($2d$-both end open)
and Va($3d$-one end open) do not exhibit such special
characteristics. Absence of this feature may be due to a small
temperature or a small system size with insufficient 
$h_2/h_1$ ratio. 

\subsection{Evaporation}
If a partially or completely filled pore is kept in vacuum, the density
inside it decreases with time. This is what we call evaporation and
investigate it for different pore geometries. In
experiments~\cite{wallacher04} one measures the vapor pressure change
inside a previously equilibrated pore subject to the 
pressure variation in the reservoir. However, we carry out the simulation
in a slightly different way by keeping the initial pore density just
above the desorption threshold.

\begin{figure}
\includegraphics[width=\linewidth]
 {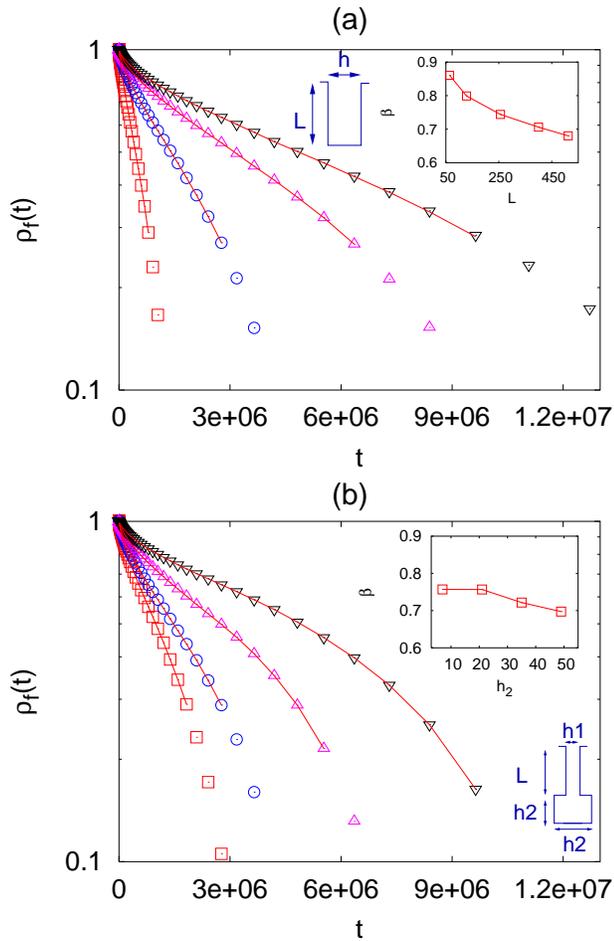}
\caption{{\bf (a):} Evaporation from simple pores of diameter $h = 7$ and
 different lengths $L=$128, 256, 400 and 512 from left to right at
 temperature $T=0.4$. Data for the filling fraction $\rho_f(t)$ are
 shown in a log-linear scale - note that the decay is initially slower
 than exponential. The full line connects the data points that can very
 well be fitted to a stretched exponential with the exponent $\beta$
 given in the {\it inset}. {\bf Inset:} The
 decay exponent $\beta$ as a function of the pore length $L$. {\bf (b):}
 $\rho_f(t)$ for ink-bottle pores of fixed length 
 $L=200$, tube diameter $h_1 = 7$ at $T = 0.4$ for different sack width
 $h_2=$7, 21, 25, 49 from left to right. The decay law is similar to the
 simple pores. {\bf
 Inset:} $\beta$ as a function of $h_2$.}
\label{fig3}
\end{figure}

To study the variation in pore density $\rho_f$ as a
function of time we fill pores completely and allow it to evaporate in
an empty reservoir(vacuum). The change in $\rho_f$ has been recorded as
a function of time $t$. Decay of the pore density can be very well
described by a stretched exponential law,

\begin{equation}
 \rho_f(t) \sim exp \left[ -(t/\tau)^{\beta} \right].
\label{eq3}
\end{equation}
The simulation is carried out in pores with both simple and ink-bottle
geometry. For simple pore we study the evaporation at temperature 
$T = $0.4,  for systems with $L =$ 128, 256, 400, 512 and $h =$7 and
finally average over 50 ensembles. The pore density $\rho_f$ as a function of
time $t$ is plotted in Fig.~\ref{fig3}{\bf (a)} in a log-linear
scale. It is noticed that a pure stretched exponential decay is found 
only above a certain value of the pore filling fraction as shown by
continuous lines.  These values of $\rho_f$ are  nothing but the ratio
between the number of surface to bulk molecules and decreases as $L$
becomes larger. The surface molecules, are attached rather strongly
($W_{wp} > W_{pp}$) to the walls and slow down the 
evaporation rate for a while but finally drop off suddenly to
zero. As the length of the system is increased, the particles deep inside
the pore requires much more trial attempts to diffuse till the open
end which effectively increases the evaporation time for longer
pores. This leads to the decay exponent $\beta$ to decrease with $L$, as
shown in the {\it inset} of {\bf (a)}.  

A similar study in case of ink-bottle pores was carried out
 with systems of fixed $L = 200$ and $h_1 = 7$ for different values of 
 $h_2 = $ 7, 21, 35, 49 as shown in Fig.~\ref{fig3}{\bf (b)}. The
 temperature is kept fixed at $T = 0.4$ 
 and data are averaged over 50 ensembles. Like the previous case, here
 also, $\rho_f$ shows a pure stretched exponential decay above a
 certain value, which becomes smaller as $h_2$ is increased. Moreover
 the effective length($L+h_2$) of the pore increase with $h_2$ results
 in the decay exponent $\beta$({\it inset} of {\bf (b)}) to drop off similarly
 to the simple pore.  
 
A further analysis(not shown) on both simple and ink-bottle pore shows that
 $\beta $ is independent of the temperature. 

\subsection{Domain evolution and Random Walks}
\begin{figure}
\includegraphics[width=\linewidth]
 {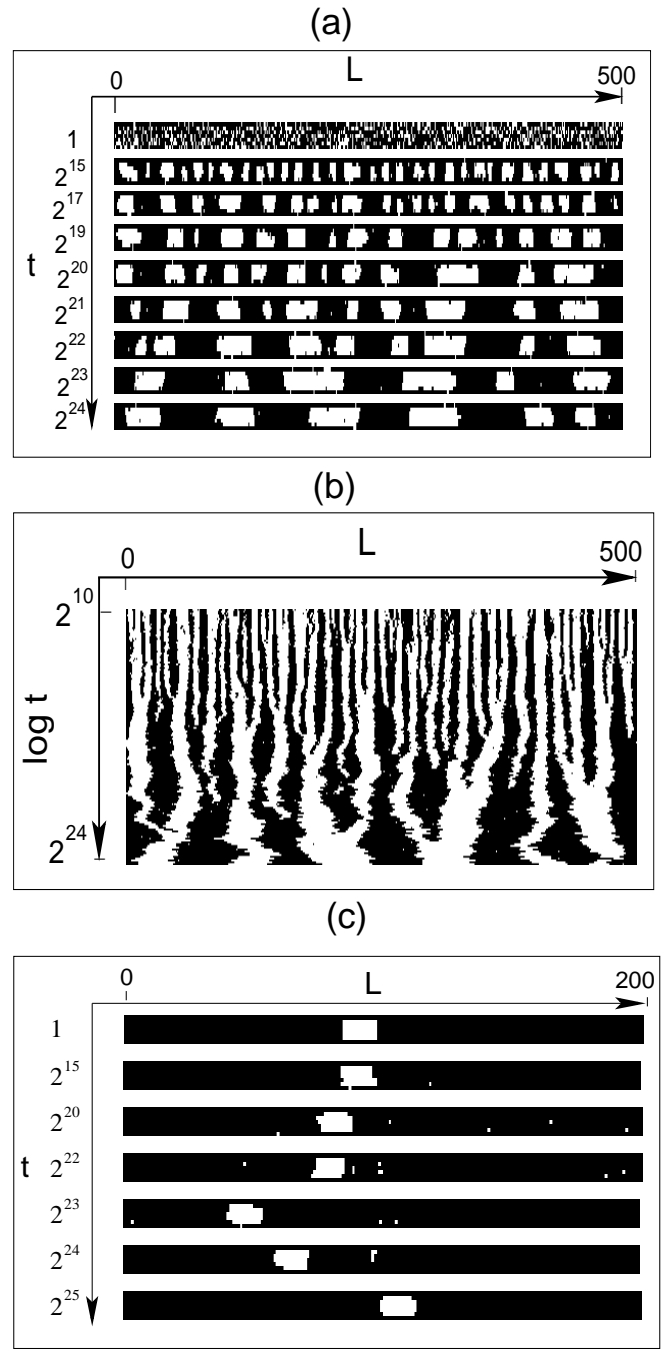}
\caption{{\bf (a):} Domain evolution in nano-pores. 
 Figures are produced by taking snap-shot for a system of length 
$L = 300$ and height $h = 7$ at temperature $T = 0.3$. {\bf (b):} 
 Axial snapshot of the domains for the same system in Fig. {\bf (a)}.
 A horizontal cut through the figure at any instant $t$ will give the
 location and size of domains/blobs along the axis of the pore. The time
 is plotted in logarithmic scale. {\bf (c):} Time evolution of a single
 rectangular blob of linear size $l_b=13$ in a pore of dimension
 $200 \times 7$ at $T=0.3$.}
\label{fig4}
\end{figure}

In this section we study the temporal evolution of domain structures
inside pores starting with a random(high temperature) initial
configuration of given density. Now we use periodic boundary
conditions between the two ends of the pore and the total number of
particle is conserved. Fig.~\ref{fig4}({\bf a}), shows the domain
structure-profile at different time steps when the system is quenched to
$T=0.3$ from $T=\infty$. Fig.~\ref{fig4}({\bf b}) is the snap-shot of
a cut along the pore axis. The black regions correspond to the
occupied sites and are called ``particle-domains'', whereas empty
white regions are termed ``blobs''.  

In the initial stage, the growth is dominated
 by nucleation and spinodal decomposition. As soon as the domain- or
 the blob-size becomes comparable to the diameter of 
 the pore, the above two mechanisms do not work any more. Transverse
 directional growth is completely stopped because of 
 the pore wall and the horizontal movement of the blobs is slowed down
 by the presence of the particle-domains.

 The snapshots in Fig.~\ref{fig4} confirm that the random motion of the
 blobs  plays an important role in the late stage of growth. A closer look to
 Fig.~\ref{fig4}(b), which is horizontally the occupancy along the
 pore-axis and vertically time in log scale, shows in the late stage, the blobs
 move to and fro along the axis of the pore and penetrate through the
 particle-domains to coalesce with the neighbors. During this process it
 also transfers a holes(vacancies, white regions) from its surface to the
 neighboring blobs.  

To elucidate the random motion, we study the time
evolution of a single blob as shown in Fig~\ref{fig3}(c). Initially it
is a perfect rectangle of linear size($l_b$) comparable to the pore
diameter($h $) and placed in the center of the pore. The temperature is 
kept fixed at $T = 0.3$. The system is allowed to evolve and the mean
square deviation $\langle x^2(t) \rangle$ of the center of 
mass(CAM) of the blob along the axis of the pore is measured at each
time step. Since the boundary of
the blob perpendicular to the wall, fluctuates very rapidly, the true COM
may not lie on the geometrical axis of the pore. In this case we trace
the actual COM and project it on to the axis. The size of the blob has
to be chosen large enough to avoid disintegration of the main blob.

Since we start with an exactly rectangular blob, a true random-walk
motion is not immediately observed. In the early stages, the blob walls, 
perpendicular to the pore axis roughens, leading to 
$\langle x^2(t)\rangle \sim t^{2/3}$ as in surface roughening described
by the KPZ-equation~\cite{kpz}. At late stages $(t>10^4)$, the blob   
performs a random walk with a blob-size dependent diffusion constant
${\mathcal D}(l_b)$,  

\begin{equation}
\langle x^2(t) \rangle = \mathcal{D}(l_b)~t.
\label{eq4}
\end{equation}

\begin{figure}
 \includegraphics[width=\linewidth]
{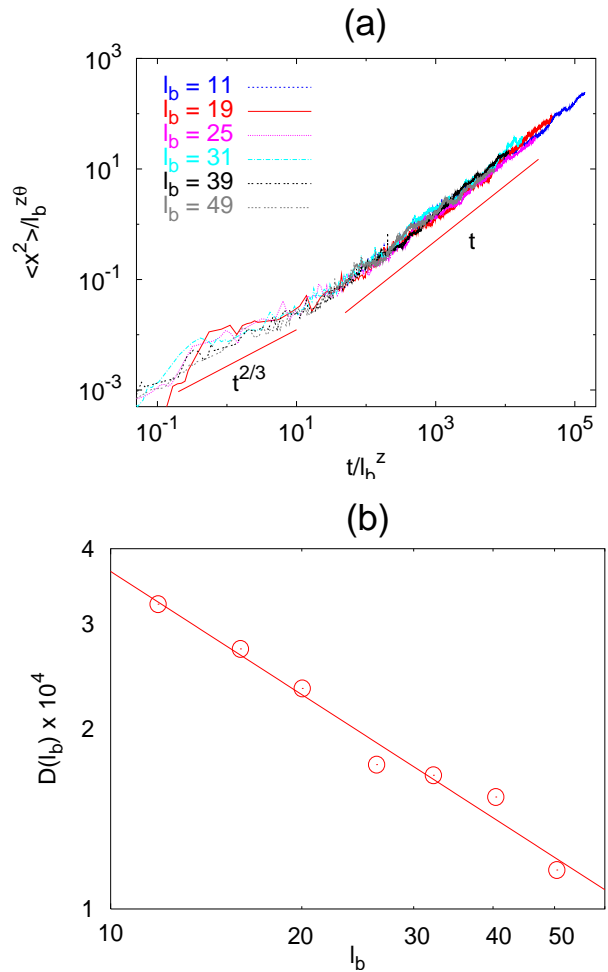}
\caption{{\bf (a):} Mean-square deviation of blobs of size $l_b=$
 11, 15, 19, 23, 27, 31, 35, 39  for a system size $L=200, h=7$ 
 and the temperature $T=0.3$. {\bf (b):} Shows the power low decay
 of the diffusion constant $\mathcal{D}(l_b)$ for different blob
 sizes. The exponent is found to be $\gamma = 0.68$. }
\label{fig5}
\end{figure}

Fig.~\ref{fig5}({$\bf a$}), shows the mean square displacement
$\langle x^2(t) \rangle $ of the COM of blobs with $l_b=$11, 15, 19,
23, 27, 31, 35, 39  as a function of time $t$. For each values of $l_b$ 
the simulation is carried out in a system with fixed $L=200$, $h=7$ at
temperature $T=0.3$ and finally averaged over 1000
ensembles. It is evident from the figure that the onset of a true
random-walk is delayed for bigger blobs. A scaling form for both 
small and large time scale regime is,
\begin{equation}
\langle x^2(l_b, t) \rangle \propto {\tau^\theta}(l_b)~f(t/\tau(l_b)
\label{eq5}
\end{equation}
with 
\begin{eqnarray}
f(x) \sim 
\left\{\begin{array}{cll}
 x^\theta & \mathrm{for}& x \ll 1 \\ 
 x        & \mathrm{for}& x \gg 1.
\end{array}\right.
\label{eq6}
\end{eqnarray}
Assuming, $\tau(l_b) \sim {l_b}^z $, from Eq.~\ref{eq5}, one can write
for large time $t \gg \tau$, 
\begin{equation}
\langle x^2(l_b, t) \rangle \propto  {l_b}^{z(\theta-1)}~t,
\label{eq7}
\end{equation}
which readily gives,
\begin{equation}
\mathcal{D}(l_b) \sim l_b^{-z(1-\theta)} = {l_b}^{-\gamma}.
\label{eq8}
\end{equation}
The scaling form in Eq.~\ref{eq5} gives quite reasonable data collapse for
$\theta = 2/3$ and $z = 2$, as shown in Fig.~\ref{fig5}(a). Substituting
these value in Eq.~\ref{eq8} we obtain the value of $\gamma = 2/3$.
Diffusion constant $\mathcal{D}$ evaluated from the asymptotic
behavior~[c. f. Fig~\ref{fig5}(a)] is plotted
in Fig~\ref{fig5}(b). From the slope of  $\mathcal{D}$ vs. $l_b$ curve
in log-log scale we estimate the $\gamma=0.68$ which agrees quite
well with the previous value 2/3, obtained from the scaling. Note that
the relation $\mathcal{D} \propto 1/l_b^{2/3}$ deviates from the naive 
expectation $\mathcal{D} \propto k_BT\eta/l_b$~\cite{siggia,wetting-liu},  
where $k_B,~T,~\eta$ are the Boltzmann' constant, temperature and
viscosity coefficient of the medium respectively.

\subsection{Correlation function and Domain growth exponent}
\begin{figure}
 \includegraphics[width=\linewidth]
{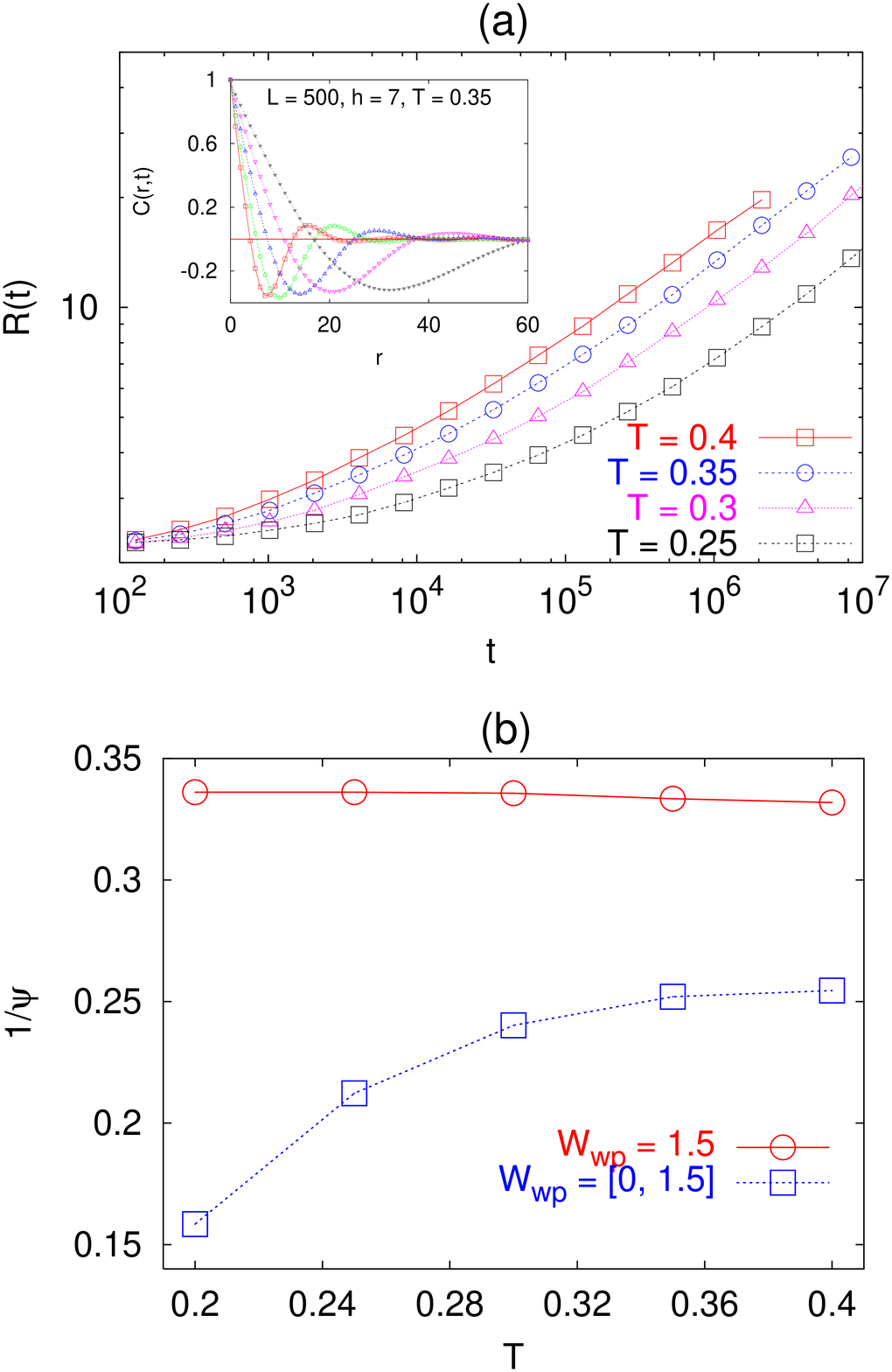}
\caption{{\bf (a):} Average domain size $R(t)$ as a function of time $t$ for
 different quenches to $T=$ 0.4, 0.35, 0.3. The system size $L = 500$,
 $h = 7$ and the density $\rho_f = 0.7$. {\bf Inset :} Correlation
 function $C(r,t)$ at various times $t$ after a quench to $T=0.35$. The
 domain size $R(t)$ is defined as the distance of the first zero from
 the origin. {\bf (b):} Growth exponent $1/\psi$ plotted against the
 temperature $T$. The upper curve, for fixed $W_{wp}$, is almost a
 constant with $T$, whereas the lower one, for random $W_{wp}$, shows a
 steep decrease for smaller values of $T$.}
\label{fig6}
\end{figure}

An alternative way to study domain growth is via the measurement of
the correlation function $C(r,t)$ along the axis of the nano-pore in a
similar fashion as in the Ising model.
\begin{equation}
C(r,t) = \langle S(0,t)S(r,t)\rangle -\langle S \rangle^2,
\label{eq13}
\end{equation}
where $S(r,t) = 2\sigma(r,t)-1$, the Lattice gas-variable, takes the
values -1, 1 for $\sigma = 0,1$ respectively. Due to this
transformation, $C(r,t)$ falls off with $r$ in an oscillatory fashion,
as is shown in the $\it inset$ of Fig.~\ref{fig6}($a$). Following
Huse~\cite{huse}, we also define the length scale or the typical domain
size $R(t)$ of the system as the position of the first zero of
$C(r,t)$. Calculating $C(r,t)$ using Eq.~\ref{eq13} we extract 
$R(t)$ by fitting the three or four points in $C(r,t)$ closest to its
first zero to a quadratic function of $r$ and defining $R(t)$ as the
value of $r$ for which the function vanishes. 
At very late stage($t > \tau(T)$), the asymptotic domain size $R(t)$
grows as a power-law,
\begin{equation}
R(t) \sim t^{1/\psi},
\label{eq14}
\end{equation}
where $1/\psi$ is the growth exponent. The onset of such a power
law regime shifts towards the early time scales as the temperature is
increased. 

We studied the domain growth for both ({\bf i}) simple and ({\bf ii})
complex pores. By ``simple'' we specify a pore with no geometrical
defects in the wall and has a fixed value of $W_{wp}$. ``Complex'' pores
are further subcatagorized into ({\bf ii}){\bf-a:} simple pore with random
$W_{wp} \in [0, 1.5]$ and ({\bf ii}){\bf-b:} pore with fixed $W_{wp}$ but
geometrical defects along the 
wall. Let us discuss about the ``simple'' pores first. The domain
evolution in this case studied with a system of length $L = 500$ and
height $h = 7$ at different temperature $T =$ 
0.25, 0.3, 0.35, 0.4. For each temperature we averaged over 300
ensembles. The average length scale or the domain size 
$R(t)$ as a function of the time $t$ is plotted in the
Fig.~\ref{fig6}({$\bf a$}). The growth exponents $1/\psi$, extracted
from this figure, for different values of temperatures, is shown by
upper curve in Fig.~\ref{fig6}({$\bf b$}). This exponent appears to be
$1/\psi \sim 1/3$ independent of the temperature. On the other
hand, for the case ({\bf ii}){\bf-a} with defects in terms of random
wall-particle coupling $W_{wp} \in [0,1.5]$, we carried out a similar
kind of study as described 
above. It is observed that the growth process is drastically slowed down by
the pinning-effect~\cite{paul04} of non-wetting sites located randomly in the
pore walls. And as a consequence the exponents are also reduced quite
significantly and found to have a temperature dependence as well, as shown in
the bottom of  Fig.~\ref{fig6}($b$).  

\begin{figure}
 \includegraphics[width=\linewidth]
{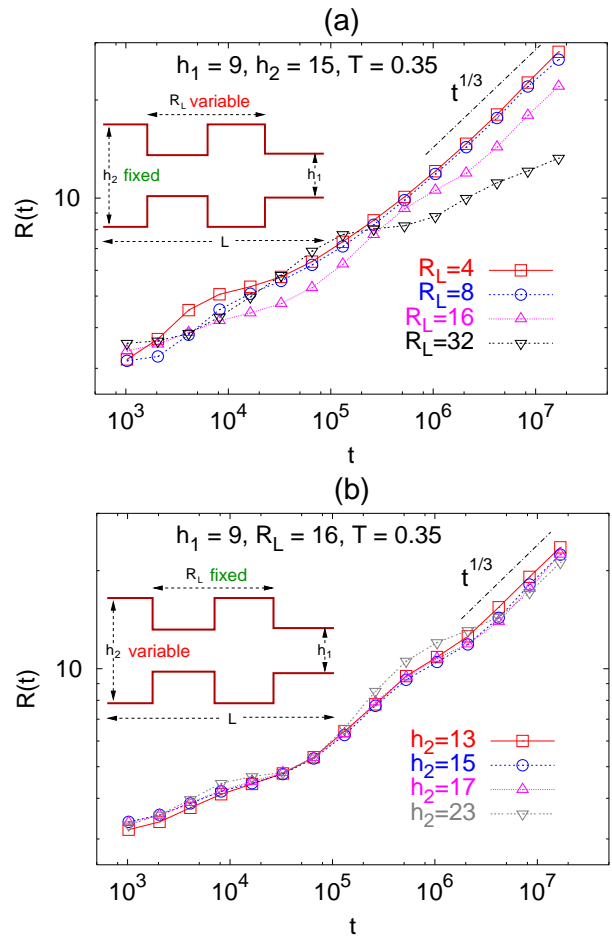}
\caption{{\bf (a):} Average domain size $R(t)$ of a defected pore-wall
 for periodicities $R_L=$ 4, 8, 16 and 32 at temperature $T=0.35$. The
 system has fixed length $L = 512$ and diameters $h_1 = 9$ and $h_2 =15$. 
 Domain growth slows down for larger periodicity. {\bf (b):} $R(t)$ for
  different pore diameters $h_2=$13, 15, 17 and 23 at fixed $L = 512$,
 $h_1=9$, $R_L=16,$ and $T = 0.35$. The domain growth remain constant
 while varying $h_2$.}
\label{fig7}
\end{figure}

It is also noticed, for the case ({\bf ii}){\bf-b}, that a periodic
structure-defect of the pore wall has a strong influence on the domain
growth. In Fig.~\ref{fig7}({$\bf a$}) we demonstrate the effect of a variable
periodicity $R_L$ for system of length $L = 512$ with two fixed pore
diameters $h_1 = 9,~h_2 = 15$. The simulations are done at a constant
temperature $T = 0.35$ and for each value of $R_L$, the data are
averaged over 300 ensembles. Domains grow quite fast in either part of
the $h_1$ or $h_2$ diameters, but finally due to the bottle-neck effect it
becomes extremely hard to transfer the particle-domains or blobs from
one side of the narrow part(diameter $h_1$) of the pore to the
other. Consequently the growth process is slowed down. But in the late
stage, when the domain size $R(t)$ becomes comparable to the
periodicity $R_L$, the blobs are in similar environment like in a
simple pore and a true power law growth sets on. As a result a longer
periodicity corresponds to a late initiation of the power law
growth. We carried out the simulation for 4 different values of
periodicity $R_L = $4, 8, 16, 32 and for the first 3 
values of $R_L$ we found a late stage growth exponent $1/\psi = 0.33$,
but for $R_L = 32$, the power law growth in not yet initiated. For such
a large periodicity one certainly has to wait extremely long.  Further, we
carried out a similar study by varying 
the pore diameter $h_2$  while keeping the periodicity $R_L$ fixed. The
numerical data are obtained with a system of $L = 512, ~h_1 = 9, ~R_L, ~T =
0.35$ and $h_2 =$13, 15, 17, 23. Finally for each value $h_2$ we average over
300 ensembles. The domain size $R(t)$ is plotted 
in Fig.~\ref{fig7}({\bf b}). In the initial stage of the growth, as described
above, the domain grows quite independently in either part of the pore
with $h_1$ and $h_2$ diameters. A true power law growth, which starts
after $R(t) \ge R_L$, is slightly delayed for larger $h_2$. As described
above, to attain a pore environment similar to the simple pore it is
necessary to equilibrate the top and bottom of the wider part($h_2$) of
the pore which takes longer as $h_2$ is increased. Finally, the
growth exponent extracted in the late stage, found to have the same
value  $1/\psi = 0.33$, as that of the pure model.\\

\section{Discussion}
Using extensive Monte Carlo simulations we have shown how hysteresis
arises in nano-pores for different pore structures. The
characteristics of the sorption branches are influenced by the shape
and size of the pore geometry. Since the hysteresis in nano-pores
occur due to diffusion, the temperature has  
to be chosen very carefully to avoid slow diffusion rate at low
temperatures or vanishing hysteresis-loop at high temperatures. Absence
of two-step desorption branch in $2d$ ink-bottle pores 
may be due to the effect of a small temperate as found in the
experiment~\cite{wallacher04}. Choice of pore
diameters for one-end open ink-bottle geometry is also an important
factor. Because of surface tension effect, a large curvature ratio
$\frac{1}{h_1}/\frac{1}{h_2}$, can cause a huge pressure difference
between the narrow and wide parts of the pore. Increase in curvature
corresponds to a decrease in vapor pressure; described by the Kelvin
equation. As an effect if the narrow part of the pore is
filled early, can block the particle to diffuse into the wider part,
unless a very high density is reached in the reservoir. 

The stretched exponential decay of the pore density $\rho_f(t)$ also agrees
qualitatively with the experiments~\cite{wallacher04}. One simple reason of
such a behavior of $\rho_f(t)$ could be due to the formation of
meniscus at the open end of the pore which eventually restricts the rate
of evaporation depending upon the radius of its curvature.

For the phase separation of binary liquids in the pore environment,
model {\bf B}~\cite{hohenberg} corresponding to dynamics with conserved order
parameter, is not well suited, as it does not account for the
transport of the order parameter by hydrodynamic flow. Modifications to
model {\bf B} by adding an ``advection'' term describes the system quite
well~\cite{bray02}.

Late stage coarsening in pore system that we studied is
effectively driven by two mechanisms:(1)Transfer of holes from the
surface of one blob to a neighboring blob and (2)the transfer of
particles from one side of the blob to the other along the pore
wall. Owing to the first mechanism a blob shrinks and the neighboring
blobs grow, whereas due to the second one a blob moves to and fro as a
whole and coalesce with another. Our numerical study for a single blob(section
III(C)), account for the contribution arises from (2) only
and gives rise to the diffusion constant ${\mathcal D} \sim l_b^{-2/3}$
of a blob of size $l_b$. The first mechanism modifies the
single-blob diffusion constant in a non-trivial way, which is difficult to
estimate. Finally, the superposition of random motion
of blobs due to mechanism (2) and the hole transfer mechanism in (1) leads to a
late-stage growth law $R(t) \sim t^{1/3}$, independent of their
individual contributions.

The domain growth  exponent $1/\psi$ is significantly decreased in
 presence of random wall-particle potential $W_{wp}$ which in practice is   
due to the effect of impurity atoms at the pore walls. The temperature
dependence of the exponent may come from a logarithmic
scaling~\cite{paul04} of barrier energy of the domains. On the other
hand geometrical defects appear to slow down the growth process only in
the early time regime. But in the late stage, as the blob size becomes
larger than the wave length of the 
geometrical disorder, which in this case is 
the periodicity of the wall-defect, the domain growth shows a pure behavior.

\begin{acknowledgements}
This work was financially supported by the Deutsche
Forschungsgemeinschaft (DFG), SFB277.
\end{acknowledgements}

\end{document}